\documentclass[11pt]{article}
\usepackage{amssymb}
\usepackage{amsmath}
\usepackage[english]{babel}
\usepackage{graphicx}

\def\Journal#1#2#3#4{{#1} {\bf #2}, #3 (#4)}

\def\RNC{\em Rivista Nuovo Cimento}

\def\NIMA{{\em Nucl. Instrum. Methods} A}
\def\NPB{{\em Nucl. Phys.} B}
\def\PLB{{\em Phys. Lett.}  B}
\def\PRL{\em Phys. Rev. Lett.}
\def\PRD{{\em Phys. Rev.} D}

\def\GaC{\em Gravitation and Cosmology}

\def\JETPL{\em JETP Lett.}

\def\CQG{\em Class. Quantum Grav.}
\def\APJ{\em Astrophys. J.}
\def\SCI{\em Science}
\def\MPLA{{\em Mod. Phys. Lett.}  A}
\def\IJTP{\em Int. J. Theor. Phys.}
\def\NJP{\em New J. of Phys.}
\def\JHEP{\em JHEP}
\def\JCAP{\em JCAP}
\def\EPHJ{\em Eur.Phys.J}
\def\BWP{\em Bled Workshops in Physics}
\def\JPCS{{\em J. Phys.:} Conf. Ser.}
\def\ARAA{\em Ann. Rev. Astron. Astrophys.}
\def\IJMPA{{\em Int. J. Mod. Phys.}  A}
\def\IJMPD{{\em Int. J. Mod. Phys.}  D}

\def\s{{\,\rm s}}
\def\g{{\,\rm g}}
\def\eV{\,{\rm eV}}
\def\keV{\,{\rm keV}}
\def\MeV{\,{\rm MeV}}
\def\GeV{\,{\rm GeV}}
\def\TeV{\,{\rm TeV}}
\def\sv{\left<\sigma v\right>}
\def\({\left(}
\def\){\right)}
\def\cm{{\,\rm cm}}
\def\K{{\,\rm K}}
\def\kpc{{\,\rm kpc}}
\def\beq{\begin{equation}}
\def\eeq{\end{equation}}
\def\bea{\begin{eqnarray}}
\def\eea{\end{eqnarray}}

\begin{document}

    \begin{center}
        \large \textbf{Towards Nuclear Physics of OHe Dark Matter}
    \end{center}

    \begin{center}
   Maxim Yu. Khlopov$^{1,2,3}$, Andrey G. Mayorov $^{1}$, Evgeny Yu.
   Soldatov $^{1}$

    \emph{$^{1}$National Research Nuclear University "Moscow Engineering Physics Institute", 115409 Moscow, Russia \\
    $^{2}$ Centre for Cosmoparticle Physics "Cosmion" 115409 Moscow, Russia \\
$^{3}$ APC laboratory 10, rue Alice Domon et L\'eonie Duquet \\75205
Paris Cedex 13, France}

    \end{center}

\medskip

\begin{abstract}

The nonbaryonic dark matter of the Universe can consist of
new stable charged particles, bound in heavy "atoms" by ordinary Coulomb
 interaction. If stable particles $O^{--}$ with charge
 -2 are in excess over their antiparticles (with charge +2), the primordial helium, formed in Big Bang
 Nucleosynthesis, captures all $O^{--}$ in
 neutral "atoms" of O-helium (OHe). Interaction with nuclei plays crucial role in the cosmological evolution of OHe and in the effects of these dark atoms as nuclear interacting dark matter. Slowed down in terrestrial matter OHe atoms cause negligible effects of nuclear recoil in underground detectors, but can experience radiative capture by nuclei. Local concentration of OHe in the matter of detectors is rapidly adjusted to the incoming flux of cosmic OHe and possess annual modulation due to Earth's orbital motion around the Sun. The potential of OHe-nucleus interaction is determined by polarization of OHe by the Coulomb and nuclear force of the approaching nucleus. Stark-like effect by the Coulomb force of nucleus makes this potential attractive at larger distances, while change of polarization by the effect of nuclear force gives rise to a potential barrier, preventing merging of nucleus with helium shell of OHe atom. The existence of the corresponding shallow well beyond the nucleus can provide the conditions, at which nuclei in the matter of DAMA/NaI and DAMA/LIBRA detectors have a few keV binding energy with OHe, corresponding to a level in this well. Annual modulation of the radiative capture rate to this level can reproduce DAMA results. The OHe hypothesis can qualitatively explain the controversy in the results of direct dark matter searches by specifics of OHe nuclear interaction with the matter of underground detectors.

\end{abstract}
\section{Introduction}
Ordinary matter around us consists of neutral atoms, in which
electrically charged nuclei are bound with electrons.
Few years ago we proposed that in the similar way the dark matter
consists of dark atoms, in which new stable charged particles are bound by ordinary Coulomb interaction (See \cite{Levels,Levels1,mpla} for review and references).
In order to avoid anomalous
isotopes overproduction, stable particles with charge -1 (and
corresponding antiparticles), as tera-particles \cite{Glashow}, should be absent \cite{Fargion:2005xz}, so that stable
negatively charged particles should have charge -2 only.

Elementary particle frames for heavy stable -2 charged species are
provided by: (a) stable "antibaryons" $\bar U \bar U \bar U$ formed
by anti-$U$ quark of fourth generation \cite{Q,I,lom,Khlopov:2006dk}
(b) AC-leptons \cite{Khlopov:2006dk,5,FKS}, predicted in the
extension \cite{5} of standard model, based on the approach of
almost-commutative geometry \cite{bookAC}.  (c) Technileptons and
anti-technibaryons \cite{KK} in the framework of walking technicolor
models (WTC) \cite{Sannino:2004qp}. (d) Finally, stable charged
clusters $\bar u_5 \bar u_5 \bar u_5$ of (anti)quarks $\bar u_5$ of
5th family can follow from the approach, unifying spins and charges
\cite{Norma}. Since all these models also predict corresponding +2
charge antiparticles, cosmological scenario should provide mechanism
of their suppression, what can naturally take place in the
asymmetric case, corresponding to excess of -2 charge species,
$O^{--}$. Then their positively charged antiparticles can
effectively annihilate in the early Universe.

If new stable species belong to non-trivial representations of
electroweak SU(2) group, sphaleron transitions at high temperatures
can provide the relationship between baryon asymmetry and excess of
-2 charge stable species, as it was demonstrated in the case of WTC
\cite{KK,KK2,unesco,iwara}.

 After it is formed
in the Standard Big Bang Nucleosynthesis (SBBN), $^4He$ screens the
$O^{--}$ charged particles in composite $(^4He^{++}O^{--})$ {\it
O-helium} ``atoms''
 \cite{I}.

In all the proposed forms of O-helium, $O^{--}$ behaves either as lepton or
as specific "heavy quark cluster" with strongly suppressed hadronic
interaction. Therefore interaction with matter of O-helium is
determined by nuclear interaction of its helium shell. These neutral primordial
nuclear interacting objects contribute to the modern dark matter
density and play the role of a nontrivial form of strongly
interacting dark matter \cite{McGuire:2001qj,Starkman}.

The qualitative picture of OHe
cosmological evolution \cite{Levels,Levels1,I,FKS,KK,unesco,Khlopov:2008rp} was recently reviewed in \cite{mpla}.
 Here we
concentrate on some open questions in the properties of O-helium dark
atoms and their interaction with matter, which are crucial for our explanation of the puzzles of dark matter
searches.

\section{O-helium interaction with nuclei}

\subsection{Structure of $O^{--}$ atoms with nuclei}
The properties of OHe interaction with matter are determined first
of all by the structure of OHe atom that follows from the general
analysis of the bound states of non-hadronic negatively charged particles $X$ with nuclei in a simple model \cite{Pospelov}, in which the nucleus is regarded as a sphere with uniform charge density. Spin dependence is not taken into account so that both the particle and nucleus are considered as scalars.

Variational
treatment of the problem \cite{Pospelov} gives for 
$$0 < a= Z Z_x \alpha A m_p R < 1$$ 
the Coulomb binding energy like in hydrogen atom, while at $$2 < a < \infty$$ for large nuclei $X$ is inside nuclear radius and the harmonic oscillator approximation is valid. Here $\alpha$ is the fine structure
constant, $R = d_o A^{1/3} \sim 1.2 A^{1/3} /(200 MeV)$ is the
nuclear radius, $Z$ is the electric charge of nucleus and $Z_x$ is
the electric charge of negatively charged particle $X$ with the mass $m_o=S_3 \TeV$. The reduced mass is $1/m= 1/(A m_p) + 1/m_o$ and for $A m_p \ll m_o$ is $m\approx A m_p$.

In the case of OHe ($Z_x=2$, $Z=2$,$A=4$) $$a = Z Z_x \alpha A m_p R \le 1,$$ what proves its
Bohr-atom-like structure, assumed in our earlier papers
\cite{I,lom,Khlopov:2006dk,KK,unesco,iwara,I2}. However, the size of
He, rotating around $O^{--}$ in this Bohr atom, turns out to be of
the order and even a bit larger than the radius $r_o$ of its Bohr
orbit, and the corresponding correction to the binding energy due to
non-point-like charge distribution in He is significant. The variational
approach \cite{Pospelov} gives in the limit of small $a$ the expression
for binding energy
\begin{equation}
    E_b(a) = (\frac{1}{2}a^2  - \frac{2}{5} a^4)/(A m_p R^2).
\end{equation}
Therefore the hydrogen-like Bohr atom binding energy of OHe 
$$E_b=\frac{1}{2} Z^2 Z_x^2 \alpha^2 A m_p = 1.6 \MeV$$ is corrected for helium final size effect as follows:
\begin{equation}
    E_b=\frac{1}{2} Z^2 Z_x^2 \alpha^2 A m_p - \frac{2}{5} Z^4 Z_x^4 \alpha^4 A^3 m_p^3 R^2 \approx 1.3 \MeV.
\end{equation}

Bohr atom like structure of OHe seems to provide a possibility to
use the results of atomic physics for description of OHe interaction
with matter. However, the situation is much more complicated. OHe
atom is similar to the hydrogen, in which electron is hundreds times
heavier, than proton, so that it is proton shell that surrounds
"electron nucleus". Nuclei that interact with such "hydrogen" would
interact first with strongly interacting "protonic" shell and such
interaction can hardly be treated in the framework of perturbation
theory. Moreover in the description of OHe interaction the account
for the finite size of He, which is even larger than the radius of
Bohr orbit, is important. One should consider, therefore, the
analysis, presented below, as only a first step approaching true
nuclear physics of OHe.

\subsection{Potential of O-helium interaction with nuclei}
The approach of \cite{Levels,Levels1,mpla} assumes the following
picture of OHe interaction with nuclei: OHe is a neutral atom in the ground state,
perturbed  by Coulomb and nuclear forces of the approaching nucleus.
The sign of OHe polarization changes with the distance: at larger distances Stark-like effect takes place - the Coulomb force of nucleus polarizes OHe so that He is put behind $O^{--}$ and nucleus is attracted by the induced dipole moment of OHe, while as soon as the perturbation by nuclear force starts to dominate the nucleus polarizes OHe in the opposite way so that He is virtually situated more close to the nucleus, resulting in a dipole Coulomb barrier for helium shell in its merging with the approaching nucleus. Correct mathematical description of this change of OHe polarization, induced by the simultaneous action of Coulomb force and strongly nonhomogeneous nuclear force needs special treatment. For the moment we use the analogy with Stark effect in the ground state of hydrogen atom and approximate the form of dipole Coulomb barrier by the Coulomb barrier in the theory of $\alpha$ decay, corrected for the Coulomb attraction of nucleus by $O^{--}$.
When helium is completely merged with the nucleus the interaction is
reduced to the oscillatory potential of $O^{--}$ with
homogeneously charged merged nucleus with the charge $Z+2$.

\begin{figure}
    \begin{center}
        \includegraphics[width=4in]{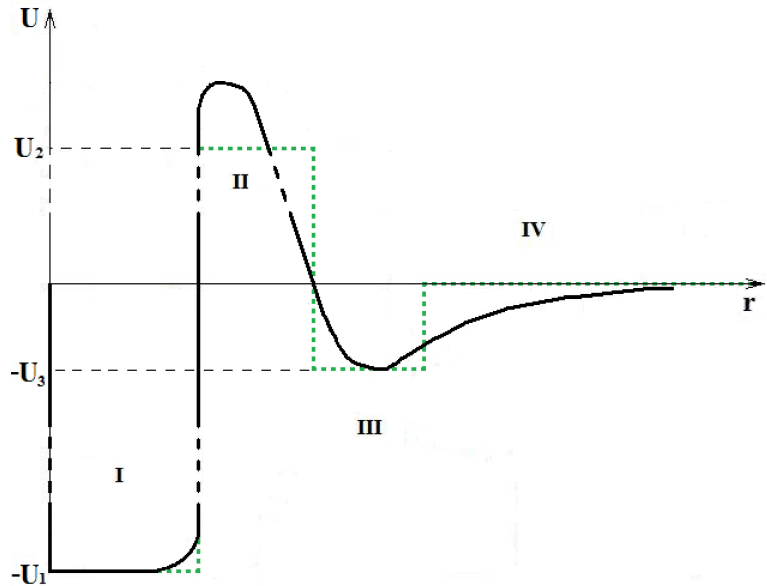}\\
        \caption{The potential of OHe-nucleus system and its rectangular well and wall approximation.}\label{pic1}
    \end{center}
\end{figure}

Therefore OHe-nucleus potential has qualitative feature, presented on Fig.~\ref{pic1} by solid line. To simplify the solution of Schrodinger equation the
potential was approximated in \cite{Levels,Levels1} by a rectangular wells and wall, shown by dashed lines on Fig.~\ref{pic1}.
The existence of potential barrier $U_2$ in region II causes suppression of reactions with transition of OHe-nucleus system to levels in the potential well $U_1$ of the region I. It results in the dominance of elastic scattering while transitions to levels in the shallow well $U_3$ (regions III-IV) should dominate in reactions of OHe-nucleus capture.

Schrodinger equation for OHe-nucleus system is reduced to the problem of relative
motion for the reduced mass $m=\frac{Am_p m_o}{Am_p+m_o}$
 in the spherically symmetric potential, presented on Fig.~\ref{pic1}.
If the mass of OHe $m_o \gg A m_p$, center of mass of
OHe-nucleus system approximately coincides with the position of
$O^{--}$ and the reduced mass is approximately equal to the mass of nucleus $Am_p$, where $A$ is its atomic weight.


Solutions of Schrodinger equation for each of the four regions,
indicated on Fig. \ref{pic1}, are given in textbooks (see
e.g.\cite{LL3}) and their sewing determines the condition, under
which a low-energy  OHe-nucleus bound state appears in the region
III.

Strictly speaking, we should deal with a
three-body problem for the system of He, nucleus and $O^{--}$ and
the correct quantum mechanical description should be based on the
cylindrical and not spherical symmetry. In the lack of the exact solution of the problem we
present here qualitative arguments for the existence and properties of OHe-nucleus bound states.

\section{OHe in the direct searches for dark matter}
\subsection{O-helium in the terrestrial matter} The evident
consequence of the O-helium dark matter is its inevitable presence
in the terrestrial matter, which appears opaque to O-helium and
stores all its in-falling flux.

After they fall down terrestrial surface, the in-falling OHe
particles are effectively slowed down due to collisions with
matter, which are dominantly elastic as follows from our description of OHe-nucleus interaction.
Then they drift, sinking down towards the center of the
Earth with velocity \cite{I}\beq V = \frac{g}{n \sigma v} \approx 80 S_3
A_{med}^{1/2} \cm/\s. \label{dif}\eeq Here $A_{med} \sim 30$ is the average
atomic weight in terrestrial surface matter, $n=2.4 \cdot 10^{24}/A$
is the number density of terrestrial atomic nuclei, $\sigma v$ is the rate
of nuclear collisions and $g=980~ \cm/\s^2$.

In underground
detectors, OHe ``atoms'' are slowed down to thermal energies and
give rise to energy transfer $\sim 2.5 \cdot 10^{-4} \eV A/S_3$, far
below the threshold for direct dark matter detection. It makes this
form of dark matter insensitive to the severe CDMS \cite{Akerib:2005kh} and XENON100 \cite{xenon} constraints. However, OHe induced processes in the matter
of underground detectors can result in observable effects. These
effects strongly
depend on the details of the OHe interaction with nuclei.

It should be noted that the nuclear cross section of the O-helium
interaction with matter escapes the severe constraints \cite{McGuire:2001qj}
on strongly interacting dark matter particles
(SIMPs) \cite{McGuire:2001qj,Starkman} imposed by the XQC experiment \cite{XQC}. Therefore, a special strategy of direct O-helium  search
is needed, as it was proposed in \cite{Belotsky:2006fa}.

Near the Earth's surface, the O-helium abundance is determined by
the equilibrium between the in-falling and down-drifting fluxes.

At a depth $L$ below the Earth's surface, the drift timescale is
$t_{dr} \sim L/V$, where $V \sim 400 S_3 \cm/\s$ is the drift velocity (\ref{dif}) and $m_o=S_3 \TeV$ is the mass of O-helium. It means that the change of the incoming flux,
caused by the motion of the Earth along its orbit, should lead at
the depth $L \sim 10^5 \cm$ to the corresponding change in the
equilibrium underground concentration of $OHe$ on the timescale
$t_{dr} \approx 2.5 \cdot 10^2 S_3^{-1}\s$.

The equilibrium concentration, which is established in the matter of
underground detectors at this timescale, is given by \cite{DMDA}
\begin{equation}
    n_{oE}=n_{oE}^{(1)}+n_{oE}^{(2)}\cdot sin(\omega (t-t_0)),
    \label{noE}
\end{equation}
where $\omega = 2\pi/T$, $T$ is the period of Earth's orbital motion around Sun and
$t_0$ is the phase.
So, there is a averaged concentration given by
\begin{equation}
    n_{oE}^{(1)}=\frac{n_o}{320S_3 A_{med}^{1/2}} V_{h}
\end{equation}
and the annual modulation of concentration characterized by the amplitude
\begin{equation}
    n_{oE}^{(2)}= \frac{n_o}{640S_3 A_{med}^{1/2}} V_E.
\end{equation}
Here $V_{h}$ is velocity (220 km/s) of Solar System in the Galaxy, $V_{E}$ is velocity (29.5 km/s) of
Earth's orbital motion around Sun and $n_{0}=3 \cdot 10^{-4} S_3^{-1} \cm^{-3}$ is the
local density of O-helium dark matter.

\subsection{OHe in the underground detectors}

The explanation \cite{Levels,DMDA} of the results of
DAMA/NaI \cite{Bernabei:2003za} and DAMA/LIBRA \cite{Bernabei:2008yi}
experiments is based on the idea that OHe,
slowed down in the matter of detector, can form a few keV bound
state with nucleus, in which OHe is situated \textbf{beyond} the
nucleus. Therefore the positive result of these experiments is
explained by annual modulation in reaction rate of radiative capture of OHe
\begin{equation}
A+(^4He^{++}O^{--}) \rightarrow [A(^4He^{++}O^{--})]+\gamma
\label{HeEAZ}
\end{equation}
by nuclei in DAMA detector.

Solution of Schrodinger equation determines the condition, under
which a low-energy  OHe-nucleus bound state appears in the shallow well of the region
III and the range of nuclear parameters was found \cite{Levels,Levels1,mpla}, at which OHe-sodium binding energy is in the interval 2-4 keV.


The rate of radiative capture of OHe by nuclei can be calculated \cite{Levels,DMDA}
with the use of the analogy with the radiative
capture of neutron by proton with the account for: i) absence of M1
transition that follows from conservation of orbital momentum and
ii) suppression of E1 transition in the case of OHe. Since OHe is
isoscalar, isovector E1 transition can take place in OHe-nucleus
system only due to effect of isospin nonconservation, which can be
measured by the factor $f = (m_n-m_p)/m_N \approx 1.4 \cdot
10^{-3}$, corresponding to the difference of mass of neutron, $m_n$,
and proton, $m_p$, relative to the mass of nucleon, $m_N$. In the
result the rate of OHe radiative capture by nucleus with atomic
number $A$ and charge $Z$ to the energy level $E$ in the medium with
temperature $T$ is given by \cite{Levels,DMDA}
\begin{equation}
    \sigma v=\frac{f \pi \alpha}{m_p^2} \frac{3}{\sqrt{2}} (\frac{Z}{A})^2 \frac{T}{\sqrt{Am_pE}}.
    \label{radcap}
\end{equation}

Formation of OHe-nucleus bound system leads to energy release of its
binding energy, detected as ionization signal.  In the context of
our approach the existence of annual modulations of this signal in
the range 2-6 keV and absence of such effect at energies above 6 keV
means that binding energy $E_{Na}$ of Na-OHe system in DAMA experiment should
not exceed 6 keV, being in the range 2-4 keV. The amplitude of
annual modulation of ionization signal can reproduce the result of DAMA/NaI and DAMA/LIBRA
experiments for $E_{Na} = 3 \keV$. The
account for energy resolution in DAMA experiments \cite{DAMAlibra}
can explain the observed energy distribution of the signal from
monochromatic photon (with $E_{Na} = 3 \keV$) emitted in OHe
radiative capture.

At the corresponding nuclear parameters there is no binding
of OHe with iodine and thallium \cite{Levels}.

It should be noted that the results of DAMA experiment exhibit also
absence of annual modulations at the energy of MeV-tens MeV. Energy
release in this range should take place, if OHe-nucleus system comes
to the deep level inside the nucleus. This transition implies
tunneling through dipole Coulomb barrier and is suppressed below the
experimental limits.

For the chosen range of nuclear parameters, reproducing the results
of DAMA/NaI and DAMA/LIBRA, the results \cite{Levels} indicate that
there are no levels in the OHe-nucleus systems for heavy nuclei. In
particular, there are no such levels in Xe, what
seem to prevent direct comparison with DAMA results in
XENON100 experiment \cite{xenon}. The existence of such level in Ge and the comparison with the results of
CDMS \cite{Akerib:2005kh} and CoGeNT \cite{cogent} experiments need special study. According to \cite{Levels} OHe should bind with O and Ca, what is of interest for interpretation of the signal, observed in CRESST-II experiment \cite{cresst}.

In the thermal equilibrium OHe capture rate is proportional to the temperature. Therefore it looks
like it is suppressed in cryogenic detectors by a factor of order
$10^{-4}$. However, for the size of cryogenic devices  less, than
few tens meters, OHe gas in them has the thermal velocity of the
surrounding terrestrial matter and this velocity dominates in the relative velocity of OHe-nucleus system.
It gives the suppression relative to room temperature
only $\sim m_A/m_o$. Then the rate of OHe radiative capture in
cryogenic detectors is given by Eq.(\ref{radcap}), in which room
temperature $T$ is multiplied by factor $m_A/m_o$. Note that in the case of $T=70\K$ in CoGeNT experiment
relative velocity is determined by the thermal velocity of germanium nuclei, what leads to enhancement relative to cryogenic germanium detectors.

\section{Discussion}

The cosmological dark matter can be formed by
stable heavy charged particles bound in neutral dark atoms by ordinary Coulomb attraction.
Analysis of the cosmological data and atomic composition of the Universe gives the constrains
on the particle charge showing that  only $-2$
charged constituents, being trapped by primordial helium
in neutral O-helium states, can avoid the problem of overproduction of the anomalous isotopes of chemical elements, which are severely constrained by observations.

This scenario can be realized in different
frameworks, in particular in Minimal Walking Technicolor model or in
the approach unifying spin and charges and contains distinct
features, by which the present explanation can be distinguished from
other recent approaches to this problem \cite{Edward} (see also
review and more references in \cite{Gelmini}).

It should be noted that O-helium, being an $\alpha$-particle with screened electric charge,
can catalyze nuclear transformations, which can influence primordial
light element abundance and cause primordial heavy element
formation. It is especially important for quantitative estimation of
role of OHe in Big Bang Nucleosynthesis and in stellar evolution.
These effects need a special detailed and complicated
study and
this work is under way. Our first steps
in the approach to OHe nuclear physics seem to support the qualitative
picture of OHe cosmological evolution described in \cite{Levels,Levels1,mpla,I,FKS,KK,unesco,Khlopov:2008rp}
and based on the dominant role of elastic
collisions in OHe interaction with baryonic matter. 

Cosmological model of O-helium dark matter
can even explain puzzles of direct dark matter searches.
The explanation is based on the mechanism of low energy binding of
OHe with nuclei. We have found \cite{Levels,Levels1} that within the uncertainty of
nuclear physics parameters there exists their range at which OHe
binding energy with sodium is equal to 4 keV and there is no such
binding with iodine and thallium. Annual modulation of the energy release in the radiative capture of OHe to this level explains the results of DAMA/NaI and DAMA/LIBRA experiments.


With the account for high sensitivity of our results to the values
of uncertain nuclear parameters and for the approximations, made in
our calculations, the presented results can be considered only as an
illustration of the possibility to explain effects in underground
detectors by OHe binding with intermediate nuclei. However, even at
the present level of our studies we can make a conclusion that
effects of such binding should strongly differ in detectors with the
content, different from NaI, and can be absent in detectors with
very light (e.g. $^3He$) and heavy nuclei (like xenon). Therefore test of results of DAMA/NaI and DAMA/LIBRA
experiments by other experimental groups can become a very
nontrivial task. Recent indications to positive result in the matter of CRESST detector \cite{cresst},
in which OHe binding is expected together with absence of signal in xenon detector \cite{xenon}, may qualitatively favor the presented approach. For the same chemical content
an order of magnitude suppression in cryogenic detectors can explain why indications to positive effect in
CoGeNT experiment \cite{cogent} can be compatible with the constraints of CDMS experiment.

An inevitable consequence of the proposed explanation is appearance
in the matter of underground detectors anomalous
superheavy isotopes, having the mass roughly by $m_o$
larger, than ordinary isotopes of the corresponding elements.

It is interesting to note that in the framework of the presented approach
positive result of experimental search for WIMPs by effect of their
nuclear recoil would be a signature for a multicomponent nature of
dark matter. Such OHe+WIMPs multicomponent dark matter scenarios
naturally follow from AC model \cite{FKS} and can be realized in
models of Walking technicolor \cite{KK2}.

The presented approach sheds new light on the physical nature of
dark matter. Specific properties of dark atoms and their
constituents are challenging for the experimental search. The
development of quantitative description of OHe interaction with
matter confronted with the experimental data will provide the
complete test of the composite dark matter model. It challenges search for stable double charged particles at accelerators and cosmic rays as direct experimental probe for charged constituents of dark atoms of dark matter.



\section {Acknowledgments}


We would like to thank Norma Mankoc-Borstnik, all the
participants of Bled Workshop and to A.S. Romaniouk for stimulating discussions.





\end{document}